\documentclass{article}

\usepackage{xcolor}
\usepackage{arxiv}

\usepackage[utf8]{inputenc} 
\usepackage[T1]{fontenc}    
\usepackage{hyperref}       
\usepackage{url}            
\usepackage{booktabs}       
\usepackage{amsfonts}       
\usepackage{nicefrac}       
\usepackage{microtype}      
\usepackage{lipsum}		    
\usepackage{graphicx}
\usepackage{natbib}
\usepackage{doi}
\usepackage{multirow} 
\usepackage{siunitx}

\usepackage{soul} 
\usepackage{amsmath}
\usepackage[percent]{overpic}
\usepackage{fontawesome5}   

\title{An End-to-End Workflow for Fin Whale Song Detection, Note Characterization, and Localization with Distributed Acoustic Sensing}
\setshorttitle{DASWhaleCalls: An End-to-End Pipeline for Fin Whale Song Analysis with DAS}

\date{} 				

\author{
{\mdseries
\textbf{D. Diego-Tortosa}$^{1}$\,\href{https://orcid.org/0000-0001-5546-3748}{\includegraphics[scale=0.035]{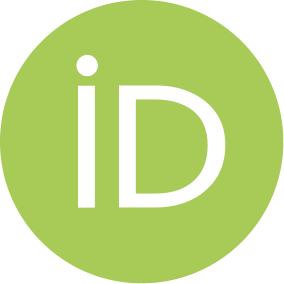}},
M. Romagosa$^{1}$\,\href{https://orcid.org/0000-0003-2781-5528}{\includegraphics[scale=0.035]{orcid.png}},
A. Ugalde$^{1}$\,\href{https://orcid.org/0000-0003-2409-2002}{\includegraphics[scale=0.035]{orcid.png}},
H. Latorre$^{1}$\,\href{https://orcid.org/0009-0003-2426-0827}{\includegraphics[scale=0.035]{orcid.png}},\\
S. Ventosa$^{1}$\,\href{https://orcid.org/0000-0002-2880-8453}{\includegraphics[scale=0.035]{orcid.png}},
J.E. García$^{2}$\,\href{https://orcid.org/0000-0002-0279-0523}{\includegraphics[scale=0.035]{orcid.png}},
A. Villaseñor$^{1}$\,\href{https://orcid.org/0000-0001-8592-4832}{\includegraphics[scale=0.035]{orcid.png}}
}\\[1ex]
$^{1}$Institut de Ciències del Mar (ICM-CSIC), Barcelona\\
$^{2}$Institut de Física Corpuscular (IFIC), Centre Mixte Universitat de València (UV-CSIC), Paterna
}

\hypersetup{
pdftitle={An End-to-End Workflow for Fin Whale anlysis with Distributed Acoustic Sensing},
pdfsubject={eess.SP, physics.geo-ph},
pdfauthor={D. Diego-Tortosa, M. Romagosa, A. Ugalde, H. Latorre, S. Ventosa, J.E. García, A. Villaseñor},
pdfkeywords={distributed acoustic sensing, ...},
}

\begin{document}
\maketitle

\begin{abstract}
Submarine fiber-optic cables instrumented with distributed acoustic sensing (DAS) provide an effective approach for large-scale monitoring of fin whales. We present an end-to-end workflow for detecting, characterizing, and localizing fin whale notes, tested on two submarine telecom cables in the Strait of Gibraltar and western Alboran Sea. The workflow applies a kurtosis-value picker adapted to narrow-band fin whale notes. Channel-wise detections are grouped into individual notes using density-based spatio-temporal clustering, cluster agglomeration, and hyperbolic fitting to reject incoherent picks. The retained clusters are characterized through temporal, spectral, and energy-related descriptors that support note-type discrimination and estimation of inter-note intervals. Relative arrival times across DAS channels are then used in a grid-search procedure to estimate candidate source locations. Evaluation against manually annotated detections from six fin whale songs yielded median pick-level precision of 0.990 and recall of 0.744, and median cluster-level precision of 0.880 and recall of 0.806. Representative applications demonstrate separation of overlapping vocalizations, characterization of type-A and type-B notes, and the inference of apparent source movement. By transforming dense DAS recordings into compact note-level bioacoustic information, the workflow provides an integrated framework for fin whale monitoring and a basis for adaptation to other synchronized acoustic receiver arrays.
\end{abstract}

\section{\label{sec:Intro} Introduction}
In recent decades, underwater acoustics has emerged as a crucial tool for studying and monitoring marine ecosystems, facilitating continuous, non-invasive observation across various spatial and temporal scales. This progress is closely associated with advancements in signal acquisition, storage, and processing technologies, which have considerably enhanced the capabilities of passive acoustic monitoring (PAM) systems. PAM enables long-term and near-real-time studies on the presence, distribution, occurrence patterns, and, in certain instances, the density of marine mammals \citep{VanParijs2009,Marques2012}. Additionally, it allows for the characterization of the underwater soundscape by assessing ambient noise levels, their temporal and seasonal variability, and their potential effects on marine fauna \citep{Havlik2022,Merchant2022}.

Hydrophones are the predominant sensors employed in underwater acoustic monitoring. These electroacoustic transducers, typically comprising one or more piezoelectric elements, function by converting acoustic pressure variations in the environment into electrical signals, which are then digitized for subsequent analysis. Their extensive use is attributed to their high sensitivity, compact size, calibration capabilities, and capacity to record signals at high sampling rates, facilitating the monitoring of a wide range of relevant frequencies. Nonetheless, their deployment in long-duration campaigns requires robust equipment capable of enduring adverse environmental conditions, as well as incurring costs related to the installation, maintenance, and retrieval of the acquisition systems \citep{Lara2020}.

These systems can be installed on autonomous platforms, which provide a flexible and cost-effective solution. However, they are constrained by energy autonomy and require equipment retrieval for data access. Alternatively, they can be installed on cabled observatories, which offer a power supply and continuous data transmission. The latter configuration supports the deployment of synchronized sensor networks, enabling not only sound detection but also the estimation of the direction of arrival and, in specific configurations, the localization and tracking of acoustic sources \citep{DiMauro2025}.

In recent years, distributed acoustic sensing (DAS) technology has emerged as a complementary approach to underwater acoustic monitoring. This technique converts fiber-optic cables into dense linear arrays of virtual sensors, providing spatially distributed measurements over tens of kilometres with sampling intervals as short as a few meters. Through optical interrogation, DAS measures dynamic axial strain or strain rate at each sampling location (hereafter referred to as a channel) providing simultaneous and synchronized observations from thousands of sensing points \citep{Hartog2017}. 

A major practical advantage of DAS is its ability to exploit existing fiber-optic infrastructures, such as submarine telecommunications cables, thereby avoiding the deployment of dedicated underwater sensor arrays and enabling continuous monitoring over large marine areas. The dense spatial sampling can also allow for the precise localization and tracking of acoustic sources. Nevertheless, the simultaneous acquisition of data from thousands of channels generates massive data volumes, requiring efficient storage, compression and processing strategies in near real time \citep{Segui2025}.

The sensitivity of DAS depends strongly on cable construction and mechanical coupling to the surrounding medium. These conditions may vary along the cable route, for example between sections that are buried, suspended, or laid on the seabed, resulting in spatial variations in the recorded signal amplitude and frequency response. In addition, DAS measures axial strain or strain rate rather than acoustic pressure, so conversion to  absolute acoustic pressure units requires characterization of the system transfer funcion. Variations in both coupling and system response restrict quantitative comparisons across different installations. Experiments involving simultaneous data acquisition with DAS and hydrophones offer a valuable opportunity to characterize DAS system responses and advance in the development of in situ calibration procedures. Several recent studies have already incorporated this type of data acquisition \citep{Matsumoto2021}. 

The recent review by \citet{Bouffaut2026pre}, along with the references cited therein, offers a comprehensive overview of the initial studies investigating the application of DAS for acoustic monitoring of marine mammals and underscores the need for methodologies specifically tailored to this monitoring approach. These pioneering studies have predominantly concentrated on baleen whales, illustrating the potential of DAS in detecting, localizing, and tracking vocalizing individuals. Among the species studied, fin whales (\textit{Balaenoptera physalus}) have garnered significant attention due to their pronounced vocal activity. Their stereotyped low-frequency vocalizations, distinguished by a unique note structure and typically lasting less than one second, render them particularly amenable to DAS-based analysis \citep{Horne2025}. These vocalizations predominantly consist of two note types: the 20 Hz pulse (type A), characterized by a downsweep from approximately 23 to 17 Hz, and the backbeat (type B), which generally spans frequencies between 18 and 20 Hz \citep{sciacca2015}. The distinctive structure of these calls makes them particularly suitable for automated long-term analysis.

Recent studies have addressed automated fin whale call detection, association, and source localization \citep{Goestchel2025,Goestchel2026,Truong2025}. The present study takes a different approach by transforming raw DAS recordings into interpretable, note-level bioacoustic information.   Its main methodological contributions are the adaptation of  the kurtosis-value picker \citep[KVP, ][]{latorre2025} algorithm to fin whale notes, and a spatio-temporal clustering and hyperbolic fitting procedure that groups channel-wise detections associated with the same vocalization while rejecting incoherent picks. The resulting note clusters are automatically characterized through the extraction of temporal, spectral, and energy-related features, supporting note-type discrimination and the calculation of inter-note intervals, and subsequently used for source localization.

\section{\label{sec:Data} Data}
\begin{figure}[t]
\centering
\begin{overpic}[width=\textwidth]{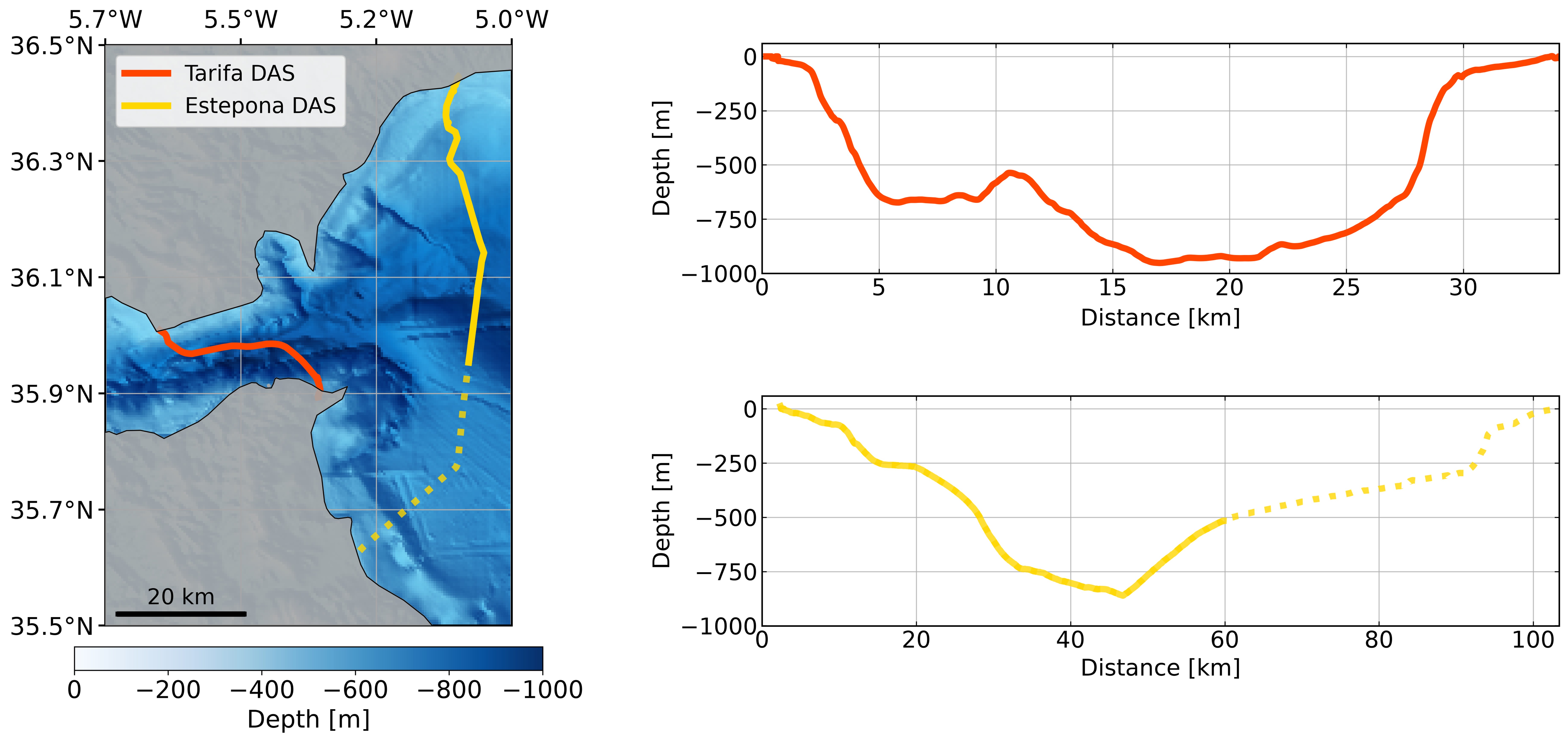}
    \put(18,-2){\small\bfseries (a)}
    \put(72.5,24){\small\bfseries (b)}
    \put(72.5,1.7){\small\bfseries (c)}
\end{overpic}
\caption{Location of the two experimental sites used in this study. \textbf{(a)} Georeferenced map showing the location of both submarine fiber-optic cables. \textbf{(b)} Bathymetric profile of the Tarifa cable. \textbf{(c)} Bathymetric profile of the Estepona cable. The solid line indicates the section interrogated by the DAS system, whereas the dotted line represents the remaining cable length beyond the interrogated range.}
\label{fig:CableMap}
\end{figure}

Two submarine DAS datasets were used to develop and evaluate the proposed processing workflow. The data were acquired on the Tarifa-Ceuta and Estepona-Tetouan fibre-optic telecommunication cables, located in the Strait of Gibraltar and the western Alboran Sea, respectively (Fig.~\ref{fig:CableMap}). A total of 34.1 km of the Tarifa-Ceuta cable and 59.8 km of the Estepona-Tetuan cable were interrogated. The Tarifa dataset was acquired between 29 June and 11 October 2022 \citep{Tarifa}, whereas the Estepona dataset was recorded between 21 September 2023 and 10 February 2024 \citep{SAFE}. Both deployments yielded extensive recordings of fin whale vocalizations, captured under varying cable geometries, propagation conditions, and signal-to-noise ratios. Previous studies have documented the presence of fin whale songs in the study area \citep{Castellote2011, Castellote2012}. The contrasting conditions represented by the two datasets provide a suitable basis for evaluating the workflow's transferability. 

Both experiments employed chirped-pulse HDAS interrogators, configured with a 10 m gauge length and 10 m channel spacing. Depending on the acquisition period, strain was recorded at either 100 or 250 Hz. Data recorded at 250 Hz were resampled to 100 Hz before analysis to standardize the workflow input sampling rate and reduce computational requirements.

\section{\label{sec:Methods} Methods}
\subsection{\label{subsec:KVP_Methods} Adaptation of KVP approach for fin whale note detection}
KVP is a multiscale, kurtosis-based algorithm originally developed for the detection and accurate picking of seismic phase arrivals \citep{latorre2025}. KVP decomposes each input trace into multiple frequency bands using a wavelet frame, computes a kurtosis-based characteristic function independently for each band, and identifies localized jumps in kurtosis associated with transient signals. Triggers occurring close in time across different bands are subsequently grouped into final picks. Because the frequency bands are analyzed independently, the resulting picks retain spectral information that can be used for signal selection and interpretation. Although  \citet{latorre2025} showed that KVP can detect fin whale vocalizations in submarine DAS recordings, the original implementation used a Ricker wavelet, selected primarily for its high temporal resolution when picking impulsive and emergent seismic arrivals. In the present study, KPV was adapted to the narrow-band, multi-cycle structure of fin whale notes by replacing the Ricker wavelet with a Morlet wavelet. 

\begin{figure}[t]
\centering
\begin{overpic}[width=\textwidth]{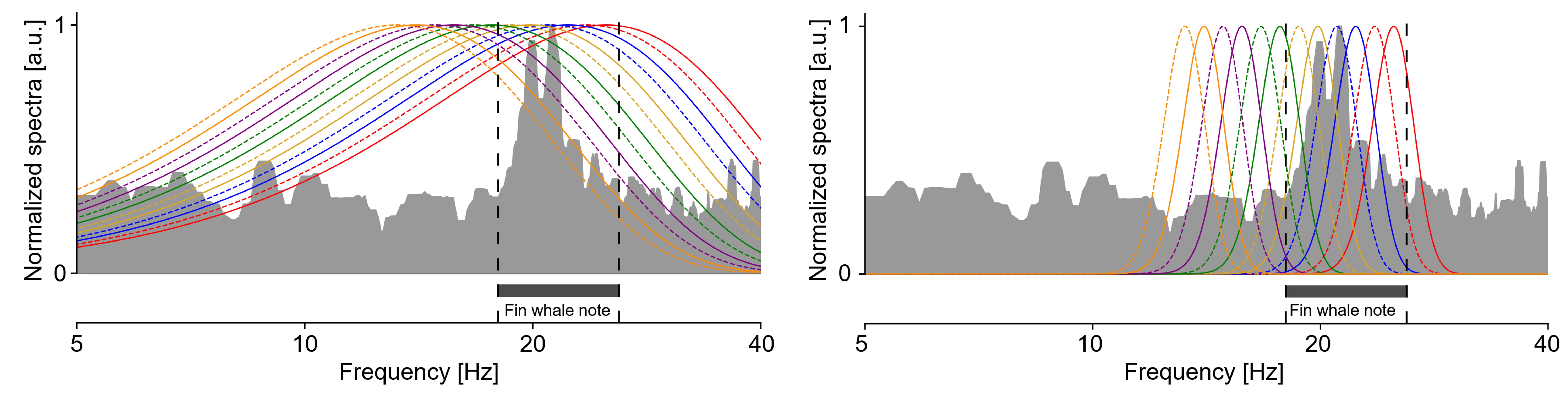}
    \put(25,-1.5){\small\bfseries (a)}
    \put(75,-1.5){\small\bfseries (b)}
\end{overpic}
\caption{Frequency response of an octave of 12 voices for both \textbf{(a)} Ricker and \textbf{(b)} Morlet wavelet families, starting at 25 Hz. The PSD of a single fin whale note is shown as a grey fill in the background, calculated by integrating the 2-D PSD around the 1500 m/s slope. All spectra are normalised to 1. A darker grey scalebar indicates the frequency band of interest for a fin whale note. Wavelet frequency responses alternate colors and line styles to help visualise their bandwidths. Frequency axis is plotted on a base 2 logarithmic scale.}
\label{fig:KVPnew}
\end{figure}

The Morlet wavelet, compared with the Ricker wavelet, provides enhanced frequency selectivity, enabling the analysis scales to isolate energy within the target frequency band more effectively while reducing out-of-band noise. Fig.~\ref{fig:KVPnew} shows the effect of wavelet choice on narrow-band signal detection. Although this modification introduces the classical trade-off between time and frequency resolution, the resulting reduction in temporal precision is not a constraint for the current application. The enhanced spectral selectivity improves the effective signal-to-noise ratio, aiding in the differentiation of signal-bearing from noise-dominated scales and reducing noise-related picks. As a result, the algorithm more reliably distinguishes fin whale notes from non-target transient signals and improves its sensitivity to weaker vocalizations. KVP is applied independently to each DAS channel, and the resulting picks are subsequently associated across channels using the spatio-temporal clustering procedure described below.

Before executing KVP, we applied a median filter to the strain records to reduce optical noise. Next, the resulting records were filtered using a zero-phase fourth-order Butterworth band-pass filter with cutoff frequencies of 10 and 40 Hz, implemented through forward--backward filtering. This relatively broad preprocessing band was selected to preserve the spectral support of the Morlet wavelet scales around the expected fin whale frequency range and to avoid truncating the response of scales near its limits. The configuration parameters used for fin whale note detection are summarized in Tab.~\ref{tab:kvp_parameters}.

\begin{table}[ht]
\centering
\caption{Morlet wavelet analysis settings and detection parameters of the adapted KVP algorithm for fin whale note detection.}
\label{tab:kvp_parameters}
\renewcommand{\arraystretch}{1.1}
\setlength{\tabcolsep}{6pt}
\begin{tabular*}{\textwidth}{@{\extracolsep{\fill}}ccccccccc@{}}
\hline
\shortstack[c]{\textbf{Max.}\\[-1pt]\textbf{freq. (Hz)}} &
\shortstack[c]{\textbf{Oct.}} &
\shortstack[c]{\textbf{Voices}\\[-1pt]\textbf{/oct.}} &
\shortstack[c]{\textbf{Center}\\[-1pt]\textbf{cyc.}} &
\shortstack[c]{\textbf{Jump}\\[-1pt]\textbf{cyc.}} &
\shortstack[c]{\textbf{Jump}\\[-1pt]\textbf{thr.}} &
\shortstack[c]{\textbf{Min.}\\[-1pt]\textbf{gap (s)}} &
\shortstack[c]{\textbf{N. bands}} &
\shortstack[c]{\textbf{Morlet}\\[-1pt]\textbf{$\omega_0$}} \\
\hline
25.0 & 1 & 10 & 150.0 & 2.0 & 2.0 & 1.0 & 2 & $2\times5.3364$ \\
\hline
\end{tabular*}
\end{table}

To exclude picks whose spectral distribution was inconsistent with that of fin whale notes, only candidates spanning no more than four frequency bands, with a minimum triggered-band frequency of at least 15 Hz and a maximum triggered-band frequency not exceeding 30 Hz, were retained for the next stage of the workflow.

\subsection{\label{subse:Fitting_Methods} Spatio-temporal clustering and hyperbolic fitting of KVP picks}
A single fin whale note generates KVP detections across neighboring DAS channels, forming a spatio-temporally coherent point pattern that approximately follows the hyperbolic moveout expected for a localized acoustic source. The objective of this stage is to group these channel-wise detections into individual clusters, each corresponding to a single fin whale note, while rejecting isolated or incoherent picks.

To this end, each KVP pick is represented by its detection time $(t)$  and its position $(d)$ along the fiber, resulting in a two-dimensional point cloud $(t,d)$. We cluster these points using the density-based spatial clustering of applications with noise algorithm, DBSCAN \citep{Ester1996}. DBSCAN was selected because detections associated with an individual fin whale note form an elongated, locally dense, and continuous pattern in the transformed time-distance space, whereas spurious KVP picks are generally more sparsely distributed. This contrast in point density enables coherent note-related detections to be grouped while isolated or weakly connected picks are treated as noise. 

DBSCAN defines clusters using a neighborhood radius, based on Euclidean distance, and a minimum number of neighboring points. Therefore, the temporal and spatial coordinates must be expressed in compatible physical units. The detection time $t$ is converted into an equivalent propagation distance $d'$ using a reference propagation velocity $c_{\mathrm{ref}}$:

\begin{equation}
\label{eq:d_prima}
d'=\frac{t}{c_{\mathrm{ref}}}
\end{equation}

where $d'$ is the time coordinate expressed as distance and $d$ is the position along the fibre. Clustering is then performed in the transformed coordinate space $(d',d)$, in which both coordinates are expressed in meters. This transformation allows the DBSCAN neighborhood radius to be specified as a physically meaningful distance, facilitating application across datasets.

Although DBSCAN identifies locally dense groups of compatible KVP detections, it may fragment the detections generated by a single fin whale note into multiple clusters. This occurs when missing KVP detections in some DAS channels interrupt the spatial continuity of the hyperbolic pattern, preventing the complete note-related structure from satisfying the density criterion required by DBSCAN.

To address this fragmentation and recover complete fin whale notes, a second agglomeration stage merges neighboring DBSCAN clusters. Each cluster is first represented by an axis-aligned bounding box in the original spatio-temporal domain $(t,d)$ which is expanded by predefined temporal and spatial margins. Clusters whose expanded bounding boxes satisfy joint proximity constraints are merged using hierarchical clustering with complete linkage, followed by a union-find procedure to enforce transitive consistency and prevent chaining effects. The resulting merged clusters represent contiguous spatio-temporal structures characterized by their temporal duration and spatial extent. However, proximity-based merging does not ensure that all detections within a cluster follow the same hyperbolic arrival pattern. A merged cluster may therefore also contain KVP detections unrelated to the corresponding fin whale note. To identify the detections that are consistent with a common acoustic arrival and reject unrelated picks, each merged cluster is subsequently fitted using the hyperbolic arrival-time model defined in Eq.~\ref{eq:hyperbolic}, following the approach proposed by \citet{Woollam2020}.

\begin{equation}
t(d)^2 = t_0^2 + \frac{(d - d_0)^2}{v_{app}^2}
\label{eq:hyperbolic}
\end{equation}

where $t(d)$ is the arrival time at distance $d$ along the DAS array, $t_0$ is the arrival time at the apex of the hyperbola, $d_0$ is position along the DAS array corresponding to the apex, and $v_{\mathrm{app}}$ is the apparent propagation velocity inferred from the hyperbolic moveout.

Direct estimation of the parameters in Eq.~\ref{eq:hyperbolic} constitutes a nonlinear optimization problem. To improve computational efficiency, a grid-search strategy was therefore adopted for $d_0$, and the solution yielding the minimum weighted root-mean-square error (RMSE) was selected.

To prioritize the earliest KVP picks arrivals, which generally provide the most reliable constraint on the hyperbola apex, a time-dependent weighting scheme was applied during fitting. The weights decreased progressively with arrival time, assigning greater influence to picks recorded near the apex, where the parameter estimation is most stable.

After the initial fit, picks with residuals exceeding a predefined threshold were classified as outliers and removed. The model was then re-estimated using the remaining picks, providing the final estimates of $d_0$, $t_0$, and $v_{\mathrm{app}}$. The quality of each fit was quantified using the mean squared error (MSE) and the coefficient of determination $R^2$. 

The output of this stage consists of a validated set of inlier KVP detections together with the estimated hyperbolic parameters describing each fin whale note. These results are subsequently used for note characterization and source localization.

\subsection{Note Characterization Features \label{subsec:NC}}
Once the validated KVP detections were obtained, the corresponding segments were retrieved from the original DAS strain records and processed independently for feature extraction. Each detection was described by its energy-weighted time and frequency centroids ($T$ and $F$); temporal and spectral spreads ($T_{\mathrm{dur}}$ and $F_{\mathrm{dur}}$); minimum and maximum frequencies ($F_{\min}$ and $F_{\max}$); mean and standard deviation of spectral energy ($E_{\mathrm{mean}}$ and $E_{\mathrm{std}}$); spectral slope and its coefficient of determination ($S$ and $S_r$); decay time ($T_{\mathrm{decay}}$); and signal-to-noise ratio ($SNR$).

This set of features was extracted following a procedure similar to the spectro-temporal pulse analysis described by \citet{Best2022}. For this analysis, the strain records were corrected for common-mode optical noise and filtered using a zero-phase sixth-order Butterworth band-pass filter with cutoff frequencies of 10--30 Hz. The data were then cropped using a 4-s temporal window starting 0.5 s before the KVP pick, resulting in the signal $s$. 

The signal $s$ was first analyzed using two time-frequency representations: a time-oriented spectrogram providing enhanced temporal resolution and a frequency-oriented spectrogram providing enhanced frequency resolution. Because temporal and frequency resolution cannot be simultaneously maximized due to the inherent time–frequency trade-off \citep{diego2025}, both spectrograms were used in the subsequent analysis.

Initially, the dominant power spectral density (PSD) peak within the 16--28 Hz frequency range, corresponding to the expected fin whale note band, was identified in the temporal-oriented spectrogram . The signal $s$ was then cropped using a 1.2 s time window centered on the time of this peak. The extracted segment, assumed to contain the fin whale note, is denoted $s_{note}$.

In the temporal domain, the extracted signal $s_{note}$ was used to compute the decay time, $T_{decay}$, and the $SNR$ according to Eq.~\ref{eq:snr}. The $SNR$ was defined as the ratio, expressed in decibels, between the maximum absolute deviation of $s_{note}$ from its median and the median absolute deviation (MAD) of $s$: 

\begin{align}
\label{eq:snr}
\mathrm{SNR}
&= 20 \log_{10} \left(
\frac{
\max \left| \mathbf{s}_{\text{note}} - \operatorname{median}(\mathbf{s}_{\text{note}}) \right|
}{
\operatorname{MAD}(\mathbf{s})
}
\right)
\end{align}

The decay time was estimated from the normalized absolute envelope of $s_{note}$, smoothed using a Savitzky–Golay filter, as the interval between the envelope maximum and the first subsequent sample at which it fell below 40\% of its peak value.

The two spectrograms were then temporally cropped to match the duration of $s_{note}$ and restricted to the 16--28 Hz frequency range. The remaining features were extracted from these representations.

First, the energy-weighted time and frequency centroids of the PSD were computed using Eqs.~\ref{eq:W} and ~\ref{eq:x_c}, yielding $T$ and $F$, respectively. The corresponding temporal and spectral spreads, $T_{dur}$ and $F_{dur}$, were defined as energy-weighted standard deviations scaled by a dimensionless factor $\sigma$, set to 1.5 in both the time and frequency dimensions to define the support of the signal, as described in Eq.~\ref{eq:x_dur}. An adaptive time–frequency mask centered at $T$ and $F$, with its temporal and spectral extents determined by $T_{dur}$ and $F_{dur}$, was used to estimate the mean spectral energy ($E_{mean}$) and its standard deviation ($E_{std}$), both expressed in dB. The minimum and maximum frequencies, $F_{min}$ and $F_{max}$, were obtained from the extrema of the spectral ridge within the temporal support of $s_{note}$.
Finally, a linear regression model was fitted to this ridge, yielding the spectral slope $S$ and its coefficient of determination $S_{r}$. The spectral ridge was defined by the frequency of maximum PSD in each time bin.

\begin{align}
\label{eq:W}
W &= \sum_{i,j} P_{ij},
\end{align}

\begin{align}
\label{eq:x_c}
X_c &= \frac{1}{W} \sum_{i,j} P_{ij}\, x_{ij},
\end{align}

\begin{align}
\label{eq:x_dur}
X_{\mathrm{dur}} &= \sigma \sqrt{
\frac{1}{W}\sum_{i,j} P_{ij}\,(x_{ij} - X_c)^2
}
\end{align}

Here $P_{ij}$ denotes the linear-scale PSD at time-frequency bin ($i$,$j$), $W$ is the total PSD weight, and $x_{ij}$ denotes either $t_j$ for the temporal descriptors or $f_i$ for the spectral descriptors. The resulting quantities are $T$ and $T_{dur}$ in the time dimension and $F$ and $F_{dur}$ in the frequency dimension.

For recordings sampled at 100~Hz, FFT lengths (\(N_{FFT}\)) of 32 and 64 were used for the time-oriented and frequency-oriented spectrograms, respectively, with an overlap of 85\%. The resulting spectrogram chacracteristics are summarized in Tab.~\ref{tab:spectrogram_parameters} \cite{diego2025}.

\begin{table}[ht]
\centering
\caption{Spectrogram parameter configurations for recordings sampled at 100~Hz. Values are rounded to two decimal places.}
\label{tab:spectrogram_parameters}
\begin{tabular*}{\textwidth}{@{\extracolsep{\fill}} ccc}
\hline
Parameter & Temporal-oriented & Frequency-oriented \\
\hline
\(N_{FFT}\) & 32 & 64 \\
Overlap [\%] & 85 & 85 \\
Temporal resolution [ms] & 48.00 & 96.00 \\
Frequency resolution [Hz] & 3.13 & 1.56 \\
Minimum valid frequency [Hz] & 6.25 & 3.13 \\
\hline
\end{tabular*}
\end{table}

The implementation additionally supports recordings with other sampling rates. In such cases, equivalent \(N_{FFT}\) values are selected automatically to preserve spectrogram characteristics comparable to those of the 100-Hz reference configuration. A set of candidate power-of-two \(N_{FFT}\) values is generated, and the corresponding temporal resolution, frequency resolution, and minimum valid frequency are computed. The search is restricted to temporal resolutions below 100~ms, frequency resolutions below 5~Hz, and minimum valid frequencies below 10~Hz. The optimal \(N_{FFT}\) for each spectrogram configuration is then selected by minimizing the relative difference from the corresponding characteristics of the 100-Hz reference spectrogram.

\subsection{Note localization}
Finally, source localization was performed for clusters containing KVP picks distributed along both branches of the fitted hyperbola. The presence of both branches provides stronger geometric constraints than a single-branch detection and helps reduce ambiguity in the estimated source position.

The proposed localization method is a simplified adaptation of the approach described by \citet{Baird2025}, designed to reduce computational complexity. It relies exclusively on the relative arrival times measured along the DAS array. Because fin whales generally remain in the upper 50 m of the water column  \citep{wat1987,Stimpert2015}, only the horizontal source position was estimated. A two-dimensional search grid was therefore generated at a fixed source depth.

The search grid was centred on the position of the earliest KVP detection in each cluster, which provides an initial approximation of the point along the cable closest to the source. The grid was defined by its horizontal extent and spatial resolution. In this study, a $6\ \mathbf{x}\ 6$ km$^2$ grid with a spacing of 25 m in both horizontal directions was used.

Only grid points located in water were retained. For each candidate source position, theoretical travel times to all DAS channels were calculated assuming a constant sound speed of 1500 m/s. These travel times were converted to relative arrival times by subtracting the minimum calculated travel time. Only channels containing an observed KVP pick were retained for comparison with the theoretical arrivals.

The observed arrival times were similarly expressed relative to the earliest detection in the cluster. For each candidate source position, the calculated and observed relative arrival times were aligned by applying a uniform temporal shift that minimized their mismatch. The shift was evaluated at increments corresponding to one sampling interval. For each tested shift, the residuals between the calculated and observed arrival times were computed. Outliers were rejected using a MAD criterion, and the optimal shift was selected as that minimizing the sum of the absolute residuals for the retained observations. The localization misfit for each candidate grid point, $\varepsilon$, is defined as the sum of the absolute residuals over the retained observations:

\begin{align}
\label{eq:P_err}
\varepsilon=\sum_{k=1}^{N_{\mathrm{in}}}\left|T_k^{\mathrm{calc}}-T_k^{\mathrm{obs}}\right|,
\end{align}

where $T_k^{\mathrm{calc}}$ and $T_k^{\mathrm{obs}}$ are the calculated and temporally shifted observed arrival times for the $k$-th inlier channel, respectively, and $N_{\mathrm{in}}$ is the number of retained observations.

A matching score $M$, was then assigned to each candidate source position. The score decreases monotonically with increasing misfit, such as the highest value corresponds to the location providing the closest agreement between calculated and observed arrival times:

\begin{align}
\label{eq:M}
M=\frac{1}{1+\varepsilon^2}.
\end{align}

The resulting score map was stored as a spatial representation of the relative agreement between each candidate source position and the observed arrival-time pattern.

\section{\label{sec:Results} Results}
Eight DAS recordings containing fin whale songs were used to develop and assess the proposed pipeline. These recordings were selected to represent a range of scenarios, including different whale positions relative to the fiber, variations in note structure and song duration, and one recording containing two simultaneously vocalizing individuals. Three recordings were obtained from the Tarifa deployment and five from the Estepona cable. Representative outputs from each stage of the workflow are presented below.

\subsection{\label{subsec:Detection_Results} Detection and clustering performance}
As an illustrative example of the proposed whale-song detection workflow, a 2-h 15-min DAS recording from the Estepona cable was selected, in which a fin whale song was visually identifiable in the KVP detections.

\begin{figure}[t]
\centering
\begin{overpic}[width=\textwidth]{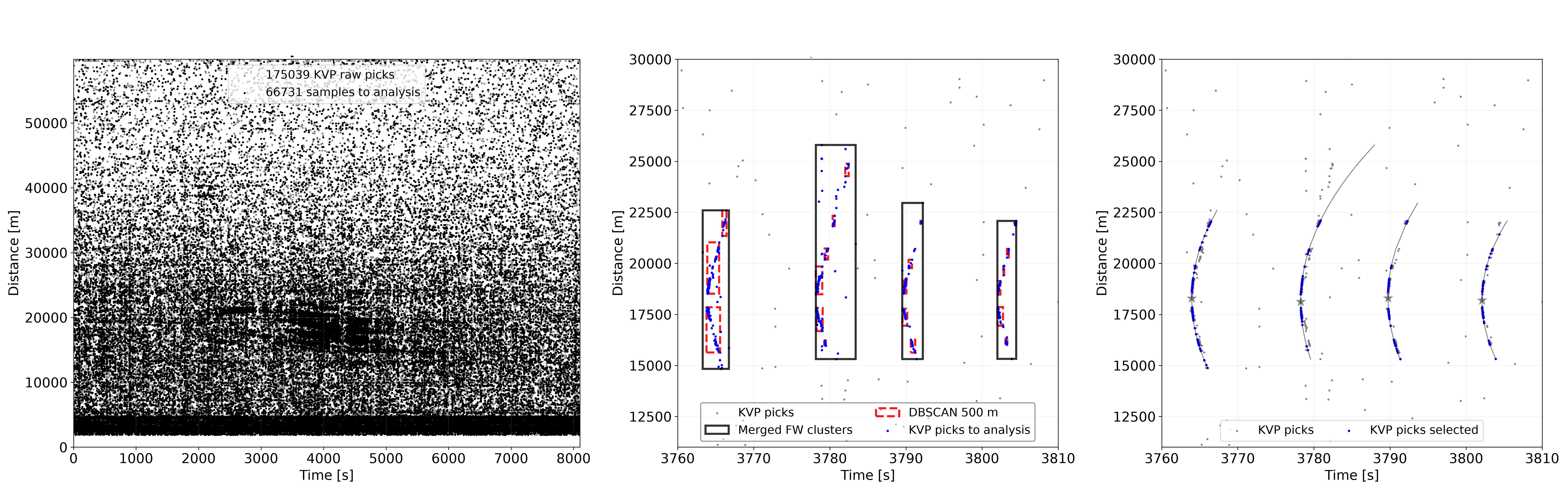}
    \put(19,-2){\small\bfseries (a)}
    \put(53.8,-2){\small\bfseries (b)}
    \put(84.5,-2){\small\bfseries (c)}
\end{overpic}
\caption{KVP picks from a 2.25-h DAS recording acquired on the Estepona cable on 13 January 2024, beginning at 06:53:00 UTC. \textbf{(a)} Raw KVP picks, with the fin whale song visible between approximately 2000 and 6000 s and between 15 and 25 km along the cable. \textbf{(b)} Enlarged view of four whale notes recorded over an approximately 7-km cable section, showing the results of the spatio-temporal clustering procedure. \textbf{(c)} KVP picks retained as inliers after hyperbolic fitting. Black stars indicate the fitted hyperbola apices.}
\label{fig:Clustering}
\end{figure}

Fig.~\ref{fig:Clustering}.a shows the 175\hspace{0.16em}039 raw KVP picks generated from the selected recording. After application of the frequency-based selection criteria described in Sec.~\ref{subsec:KVP_Methods}, 66\hspace{0.16em}731 KVP picks were retained for further analysis, corresponding to a reduction of approximately 62\%. A high density of detections remained within the first 5 km from the interrogator, where the background noise was strongest. Fig.~\ref{fig:Clustering}.b shows an enlarged view of four fin whale notes after the spatio-temporal clustering stage. This example illustrates the role of the cluster-merging procedure described in Sec.~\ref{subse:Fitting_Methods}, which consolidates adjacent clusters associated with the same acoustic event. Fig.~\ref{fig:Clustering}.c shows the corresponding hyperbolic fits and the subset of KVP picks retained as inliers for each note, illustrating the ability of the procedure to isolate detections associated with individual fin whale notes.

The detection stage was quantitatively evaluated using six fin whale song recordings for which KVP picks associated with fin whale vocalizations were manually annotated. The automatic procedure, comprising spatio-temporal clustering and hyperbolic fitting as described in Sec.~\ref{subsec:KVP_Methods}, was evaluated at both the pick and cluster levels. At the pick level, the procedure achieved a median precision of $0.990$ (IQR: $0.984$--$0.994$) and a median recall of $0.744$ (IQR: $0.575$--$0.879$). At the cluster level, the median precision was $0.880$ (IQR: $0.818$--$0.922$), whereas the median recall was $0.806$ (IQR: $0.751$--$0.885$). Thus, pick-level selection was more precise, whereas cluster-level detection achieved higher recall.

\subsection{\label{subsec:Characterization_Results} Characterization of representative fin whale notes}

Following the identification of candidate fin whale notes, the features described in Sec.~\ref{subsec:NC} were estimated from the temporal waveform, and the time- and frequency-oriented spectrograms.

\begin{figure}[t]
\centering
\begin{overpic}[width=\textwidth]{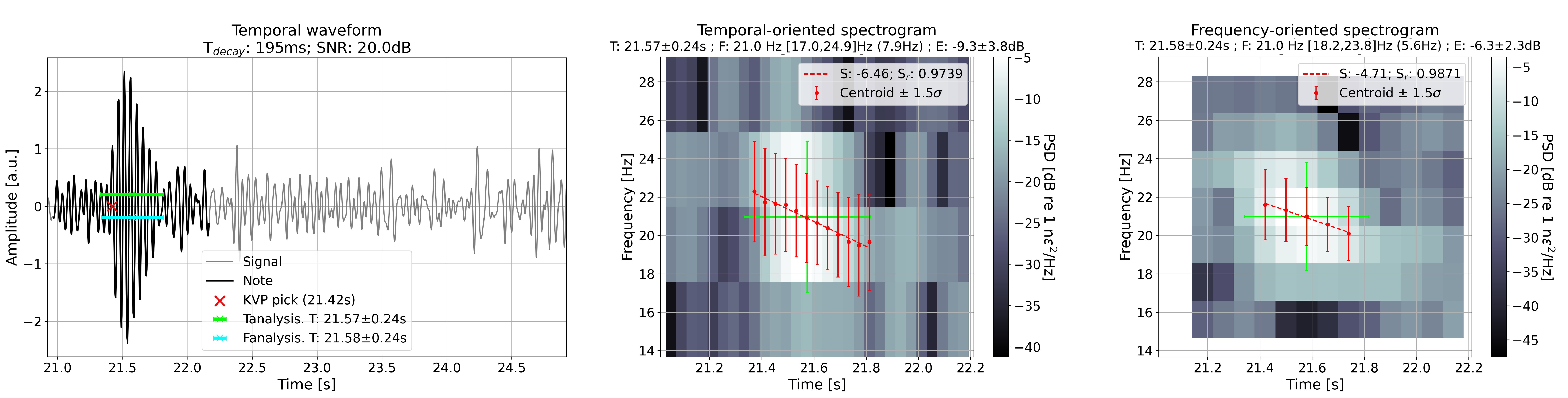}
    \put(18.3,-3){\small\bfseries (a)}
    \put(50.7,-3){\small\bfseries (b)}
    \put(82.3,-3){\small\bfseries (c)}
\end{overpic}
\caption{Characterization of a type A fin whale note recorded on the Estepona cable on 14 October 2023 at 01:41:25 UTC at channel 1739. Features extracted from \textbf{(a)} the temporal waveform, \textbf{(b)} the time-oriented spectrogram, and \textbf{(c)} the frequency-oriented spectrogram are indicated in the corresponding titles and annotations.}
\label{fig:noteA}
\end{figure}

\begin{figure}[t]
\centering
\begin{overpic}[width=\textwidth]{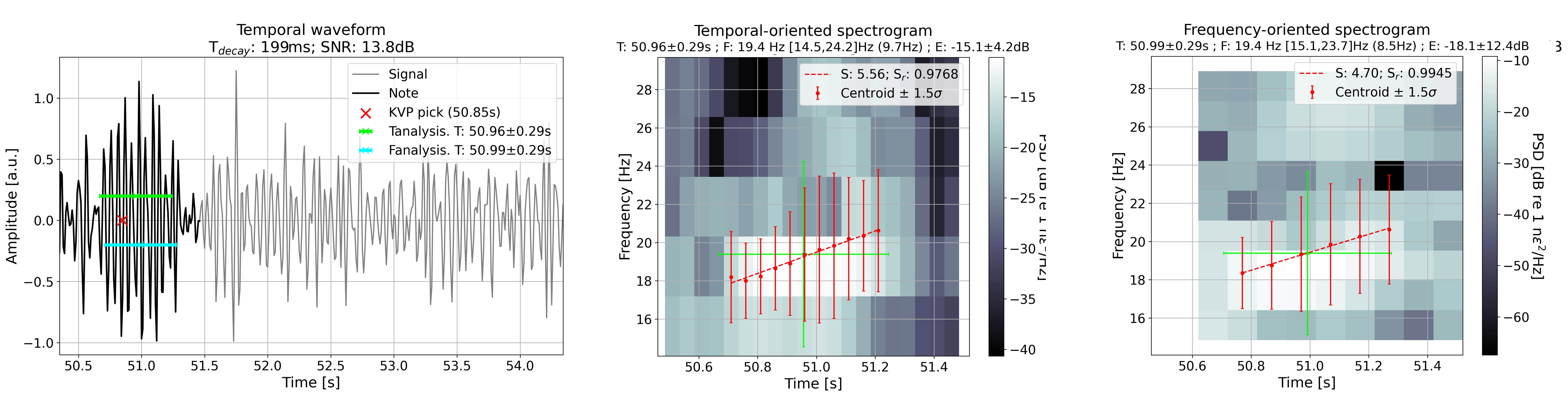}
    \put(18.3,-3){\small\bfseries (a)}
    \put(50.7,-3){\small\bfseries (b)}
    \put(82.3,-3){\small\bfseries (c)}
\end{overpic}
\caption{Characterization of a type B fin whale note recorded on the Estepona cable on 10 January 2024 at 00:10:34 UTC at channel 3121. Features extracted from the \textbf{(a)} the temporal waveform, \textbf{(b)} the time-oriented spectrogram, and \textbf{(c)} the frequency-oriented spectrogram are indicated in the corresponding titles and annotations.}
\label{fig:noteB}
\end{figure}

Figs.~\ref{fig:noteA} and ~\ref{fig:noteB} show representative type A and type B fin whale notes, respectively, extracted from DAS recordings acquired on the Estepona cable. Together, these figures illustrate the feature extraction procedure for both note types. The temporal, spectral, and energy-related descrptors are indicated in the waveform and spectrogram representations.

\subsection{\label{subsec:Localization_Results} Localization of a fin whale song}

\begin{figure}[t]
\centering
\begin{overpic}[width=\textwidth]{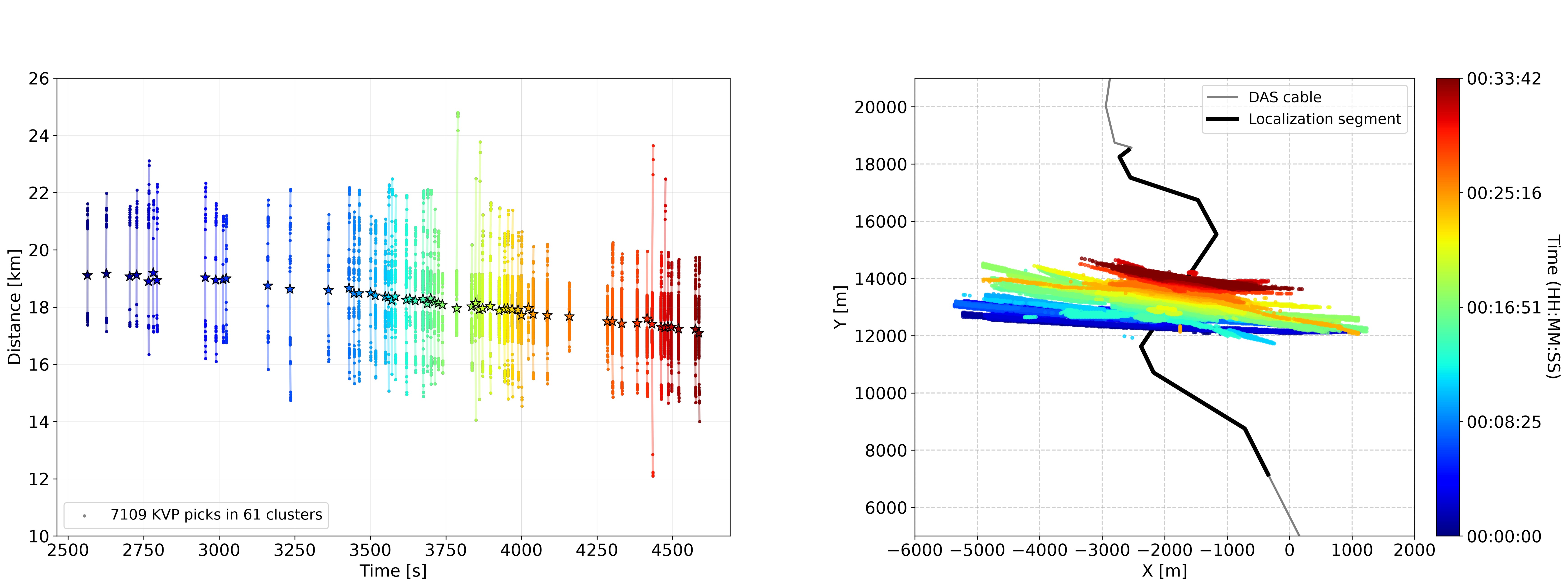}
    \put(23.5,-2){\small\bfseries (a)}
    \put(73,-2){\small\bfseries (b)}
\end{overpic}
\caption{Representative localization results for a fin whale song recorded on the Estepona cable beginning on 13 January 2024 at 06:54:30 UTC. \textbf{(a)} Selected sequence of 61 consecutive note clusters spanning 33~min~42~s. \textbf{(b)} Contours enclosing grid points with matching scores at or above the 99$^{\mathrm{th}}$ percentile of each cluster-specific score map, colored according to the apex time. Coordinates are expressed in a local Cartesian reference frame centered at the midpoint of the DAS cable.}
\label{fig:win_0031}
\end{figure}

To illustrate the localization stage, a sequence of consecutive note clusters was chosen to show the temporal evolution of the estimated source position over a complete fin whale song. Fig.~\ref{fig:win_0031}.a shows the selected clusters, which satisfied the following localization quality criteria: at least five KVP picks on each branch of the fitted hyperbola, consistent apex positions $d_0$ between successive clusters, and spatially compact high-score regions in the matching-score maps. The sequence comprised 61 consecutive clusters spanning 33 min 42 s.

Fig.~\ref{fig:win_0031}.b shows contours corresponding to the 99$^{\mathrm{th}}$ percentile of the matching score map, $M$, for each cluster, colored according to the apex time $t_0$. The temporal progression of these contours illustrates the evolution of the estimated source position during the vocalization sequence and allows the apparent direction of movement to be inferred. The localization maps retain the expected bilateral ambiguity with respect to the cable. Results are shown in a local Cartesian coordinate system whose origin is defined at the midpoint of the DAS cable  at longitude $-5.08078^\circ$ and latitude $36.19937^\circ$.

\section{\label{sec:Discussion} Discussion}
The proposed pipeline was developed and evaluated using DAS measurements, but several of its processing stages are not inherently restricted to fiber-optic sensing. KVP operates on individual time series, and the note-characterization procedure is similarly based on single-receiver waveforms and spectrograms. The association and localization stages, however, require synchronized recordings from multiple receivers with known positions and sufficient spatial sampling to resolve coherent arrival-time patterns. With appropriate adaptation of these stages, the workflow could therefore be applied to other array-based acoustic systems, including conventional hydrophone arrays, cabled marine observatories, or ocean bottom seismometer (OBS) networks.

\subsection{Strengths and limitations of the detection workflow} 
Unlike many recent detection approaches, KVP uses conventional signal-processing methods rather than supervised machine learning and therefore does not requiere previously labelled data or model training. This is particularly advantageous when annotated datasets are scarce or unavailable. Conventional signal-processing methods may also offer lower computational requirements and greater interpretability than many data-driven models.

A notable strength of KVP is its capacity to detect signals exhibiting the time–frequency characteristics represented by the selected wavelet scales, providing flexibility with respect to the specific signal source. When applied independently to the large number of channels in a DAS dataset, however, this broad sensitivity can also produce numerous non-target detections arising from spatially variable noise and cable response. An additional post-processing stage is therefore required to distinguish coherent note-related detections from isolated or incoherent picks. In the present workflow, this was achieved through spatio-temporal clustering and hyperbolic fitting, which reduced non-target detections while reatining coherent arrival patterns associated with fin whale notes.

\begin{figure}[t]
\centering
\includegraphics[width=\textwidth]{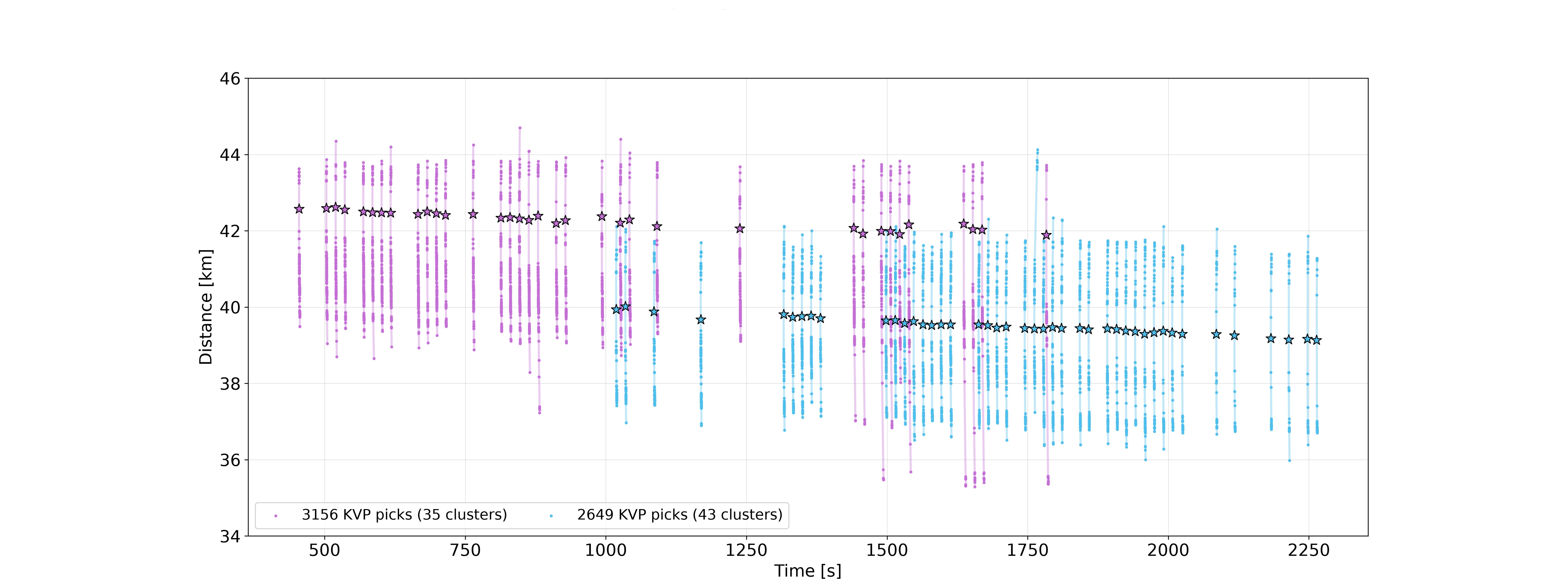}
\caption{Representative sequence of 78 consecutive clusters corresponding to two temporally interleaved fin whale songs recorded on the Estepona cable on 15 January 2024, beginning at 09:03:02 UTC. The sequence comprises an approximately 22-min song segment and a second approximately 21-min segment, with hyperbola apices located in distinct sections of the cable, consistent with two simultaneously vocalizing individuals.}
\label{fig:win_0033}
\end{figure}

An additional capability of the proposed detection workflow is the identification of simultaneous vocalizations from spatially separated sources. Following spatio-temporal clustering and hyperbolic fitting, the example shown in Fig.~\ref{fig:win_0033} contains two temporally interleaved sequences of note clusters. One sequence has hyperbola apices located between 42 and 43 km along the cable, whereas the second has apices between 39 and 40 km. The example therefore shows that overlapping note sequences can be separated when their arrival patterns and apex positions differ sufficiently.

Nevertheless, although both song sequences can be distinguished at the detection stage, the workflow does not inherently assign individual note clusters to specific whales or reconstruct continuous movement trajectories. Such an analysis would require a subsequent stage combining source localization with track association.

\subsection{Interpretation and applications of note-level features}
The features computed during the note-characterization stage are derived from two complementary time–frequency representations: a time-oriented spectrogram with enhanced temporal resolution and a frequency-oriented spectrogram with enhanced frequency resolution. Although some descriptors are estimated from both representations, the resulting values provide complementary information, particularly for distant or low-SNR detections for which reduced DAS channel sensitivity may compromise the clarity of the spectral pattern. Agreement between corresponding features obtained from both representations can therefore provide an internal consistency check, whereas substantial discrepancies may indicate an unreliable characterization.

Instantaneous frequency estimation, following  the approach proposed by \citet{Horne2025}, was also investigated for calculating $F$, $F_{min}$, and $F_{max}$. For the present DAS dataset, however, it did not provide clearer or more stable estimates of these frequency-related features than the spectrogram-based approach.

The extraction of note parameters establishes a direct link between DAS detections and biologically significant acoustic information. The features are initially estimated at the channel level each inlier detection and should subsequently be interpreted at the cluster level, because all detections within a cluster are assumed to correspond to the same note. Channel-wise estimates may vary because of differences in $SNR$ and channel sensitivity. Accordingly, note interpretation should be based on the ensemble of channel-wise estimates within each cluster rather than on individual channel observations, with higher-SNR channels generally providing the most reliable characterization.

For high-quality detections with $SNR$ values greater than 8 dB, such as those shown in Figs.~\ref{fig:noteA} and ~\ref{fig:noteB}, frequency- and slope-related features may support discrimination between type-A and type-B fin whale notes. This capability could facilitate the estimation of inter-note intervals (INIs) separately for AA, AB, BB, and BA note sequences, providing information on fin whale song structure and temporal patterns \citep{Romagosa2024}. When two consecutive fin whale notes are detected on the same channel, their arrival-time difference provides a channel-specific interval estimate, with multiple shared channels providing replicate measurements. When consecutive notes are not captured on any common channel, the interval can instead be estimated from the difference between the apex times, $t_0$, of their fitted hyperbolas.

\subsection{Source localization and movement inference}
The localization strategy described in Sec.~\ref{subsec:Localization_Results} provides a framework for analyzing the movements of singing fin whales. The method on the selection of high-quality note clusters and the use of high-percentile contours of the  matching-score maps to base the localization analysis on the most reliable detections.  KVP pick times are used instead of feature $T$, because the former show greater stability across channels, whereas $T$ exhibits increased variability with decreasing signal quality and increasing source-to-channel distance.

\begin{figure}[t]
\centering
\begin{overpic}[width=\textwidth]{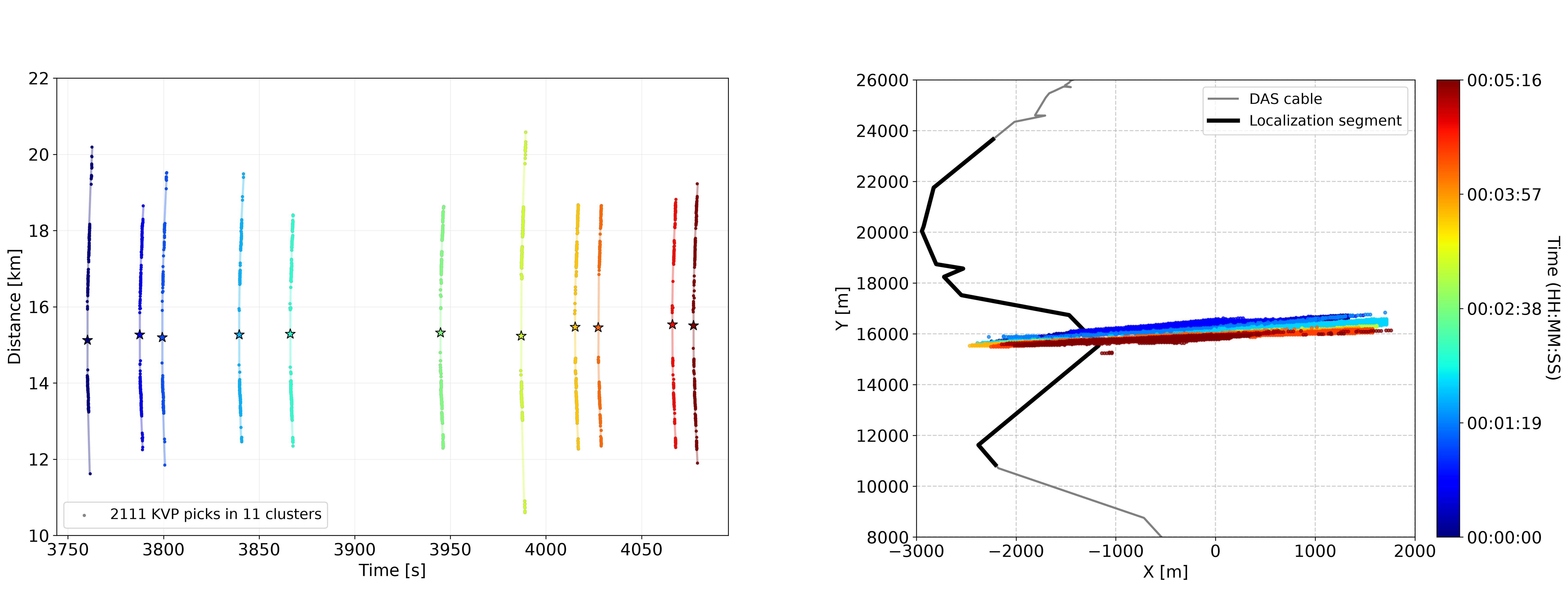}
    \put(23.5,-2){\small\bfseries (a)}
    \put(73,-2){\small\bfseries (b)}
\end{overpic}
\caption{Representative localization results for a fin whale song recorded on the Estepona cable on 14 October 2023, beginning at 00:34:12 UTC. \textbf{(a)} Selected sequence of 11 consecutive note clusters spanning 5~min~16~s song. \textbf{(b)} Contours enclosing grid points with matching scores at or above 99$^{\mathrm{th}}$ percentile os each cluster-specific score map. Coordinates are expressed in a local Cartesian reference frame centered at the midpoint of the DAS cable.}
\label{fig:win_0006}
\end{figure}

Although the matching-score maps generated for each note require additional post-processing to reconstruct whale trajectories, they already provide relevant information in their current form. As illustrated in Sec.~\ref{subsec:Localization_Results},  the sequential localization of notes enables the general direction relative to the cable of movement during a song to be inferred, thereby facilitating the study of movement patterns adirectly from the workflow output. The localization results remain affected by the inherent bilateral ambiguity associated with a single, predominantly linear DAS cable. Nevertheless, the inferred movements correspond to biologically plausible swimming speeds. For the trajectory depicted in Fig.~\ref{fig:win_0031}.b, the estimated displacement was approximately 1861~m over the 33~min~42~s duration of the song, corresponding to an apparent swimming speed of 3.31~km/h. Similarly, the example in Fig.~\ref{fig:win_0006} covered an estimated displacement of approximately 570~m in 5~min~16~s, corresponding to an apparent swimming speed of 6.48~km/h. Both estimates are below the 7~km/h swimming speeds typically reported for singing fin whales \citep{Clark2019}. 

The latter example also underscores the influence of cable geometry on localization ambiguity. The reduced linearity of the cable segment in Fig.~\ref{fig:win_0006} results in a more spatially constrained position estimate than that obtained along the predominantly straight section shown in Fig.~\ref{fig:win_0031}. This observation suggests that DAS deployments incorporating cable sections with greater directional diversity, such as curved or zig-zag layouts, or the joint interrogation of nearby cables \citep{Rorstadbotnen2023}, could reduce the inherent localization ambiguity. Such acquisition geometries should therefore be considered when designing future DAS experiments aimed at localizing and tracking marine mammals.

Finally, the matching score $M$ provides a relative measure of the agreement between the calculated and observed arrival-time patterns at each candidate source position. Rather than returning a single position, the resulting score map shows the spatial distribution of candidate locations producing comparatively good fits. It therefore provides a useful representation of localization ambiguity.

\section{\label{sec:Conclusion} Conclusion}
This work presents an automated workflow for the detection, characterization, and localization of fin whale vocalizations from raw DAS data. The proposed pipeline was applied to two DAS datasets acquired under different cable geometries and recording conditions. Its detection stage was quantitatively evaluated using manually annotated fin whale songs, while its operation does not require supervised training. The combination of KVP detection, spatio-temporal clustering, hyperbolic fitting, note characterization, and matching-score localization provides an integrated framework for transforming raw DAS measurements into interpretable bioacoustic information.

Beyond automatic detection, the proposed workflow extracts note-level acoustic features that may support discrimination between type-A and type-B fin whale notes and estimates spatial matching-score maps for the detected sources. These outputs provide the information required to estimate the INIs for AA, AB, BA, and BB note sequences and to infer apparent swimming direction during individual songs. The score maps additionally represent the relative spatial agreement between candidate source positions and the observed arrival-time patterns, although they should not be interpreted as calibrated localization uncertainty. Overall, the workflow provides a scalable methodology for investigating marine mammal monitoring using DAS and could, with appropriate adaptation, be applied to other synchronized acoustic receiver arrays.

\paragraph{\faGithub~Source code}
An open-source code with the proposed pipeline is available on GitHub: \url{https://github.com/b-csi/DASWhaleCalls}.
 
\paragraph{Acknowledgments.}
The contribution of Dr. Diego-Tortosa was supported by the \textit{ALLIES Cofund Program} which received funding from the European Union's \mbox{Horizon-MSCA-2022-COFUND-01} research and innovation program under the Marie Skłodowska-Curie grant agreement \mbox{No~101126626}. 
The contribution of Dra. Romagosa was supported by a Juan de la Cierva postdoctoral contract (\mbox{JDC2023-050731-I}).
Funding for this project was provided through the "Severo Ochoa" grant for excellence \mbox{CEX2024-001494-S} funded by AEI \mbox{10.13039/501100011033}; the European Union Next Generation EU/PRTR Program under project PSI (ref. \mbox{PLEC2021-007875}); the MCIN/AEI under project FERMAT (ref. PID \mbox{2024-162301OB-C21}); and the CSIC contract SAFE with Telxius Cable \mbox{España, S. L.} (ref. 20233069). Telxius, and GTD Group facilitated fibre-optic cable access, and Aragon Photonics provided the HDAS interrogator used in all experiments. This work is a contribution of the Barcelona-CSI that is a Grup de Recerca 2021 \mbox{SGR 00429} of Generalitat de Catalunya.

\bibliographystyle{plainnat}
\bibliography{references}

@article{Romagosa2024,
  title = {Fin whale song evolution in the {North Atlantic}},
  volume = {13},
  ISSN = {2050-084X},
  DOI = {10.7554/elife.83750},
  journal = {{eLife}},
  publisher = {eLife Sciences Publications,  Ltd},
  author = {Romagosa, M. and Nieukirk, S. and Cascão, I. and Marques, T. A. and Dziak, R. and Royer, J.-Y. and O'Brien, J. and Mellinger, D. K. and Pereira, A. and Ugalde, A. and Papale, E. and Aniceto, S. and Buscaino, G. and Rasmussen, M. and Matias, L. and Prieto, R. and Silva, M. A.},
  year = {2024},
  month = jan 
}

@Article{Goestchel2026,
  author    = {Goestchel, Quentin and Wilcock, William S. D. and Abadi, Shima},
  journal   = {The Journal of the Acoustical Society of America},
  title     = {Automated association of fin whale calls for localization using distributed acoustic sensing},
  year      = {2026},
  issn      = {1520-8524},
  month     = Jul,
  number    = {1},
  pages     = {135--146},
  volume    = {160},
  doi       = {10.1121/10.0044257},
  publisher = {Acoustical Society of America (ASA)},
}

@Article{Truong2025,
  author    = {Truong, Khanh and Eidsvik, Jo and Rørstadbotnen, Robin Andre and Morten, Jan Petter and Andres, Laurine and Sladen, Anthony},
  journal   = {The Journal of the Acoustical Society of America},
  title     = {Automated detection of fin whales with distributed acoustic sensing in the Arctic and Mediterranean},
  year      = {2025},
  issn      = {1520-8524},
  month     = dec,
  number    = {6},
  pages     = {5057--5074},
  volume    = {158},
  doi       = {10.1121/10.0041855},
  publisher = {Acoustical Society of America (ASA)},
}

@article{Stimpert2015,
  title = {{Sound production and associated behavior of tagged fin whales (Balaenoptera physalus) in the Southern California Bight}},
  volume = {3},
  ISSN = {2050-3385},
  DOI = {10.1186/s40317-015-0058-3},
  number = {1},
  journal = {Animal Biotelemetry},
  publisher = {Springer Science and Business Media LLC},
  author = {Stimpert,  Alison K and DeRuiter,  Stacy L and Falcone,  Erin A and Joseph,  John and Douglas,  Annie B and Moretti,  David J and Friedlaender,  Ari S and Calambokidis,  John and Gailey,  Glenn and Tyack,  Peter L and Goldbogen,  Jeremy A},
  year = {2015},
  month = Aug 
}

@Article{Goestchel2025,
  author    = {Goestchel, Quentin and Wilcock, William S. D. and Abadi, Shima},
  journal   = {The Journal of the Acoustical Society of America},
  title     = {Enhancing fin whale vocalizations in distributed acoustic sensing data},
  year      = {2025},
  issn      = {1520-8524},
  month     = May,
  number    = {5},
  pages     = {3655--3666},
  volume    = {157},
  doi       = {10.1121/10.0036696},
  publisher = {Acoustical Society of America (ASA)},
}

@article{Matsumoto2021,
  title = {Detection of hydroacoustic signals on a fiber-optic submarine cable},
  volume = {11},
  ISSN = {2045-2322},
  DOI = {10.1038/s41598-021-82093-8},
  number = {1},
  journal = {Scientific Reports},
  publisher = {Springer Science and Business Media LLC},
  author = {Matsumoto, Hiroyuki and
          Araki, Eiichiro and
          Kimura, Toshinori and
          Fujie, Gou and
          Shiraishi, Kazuya and
          Tonegawa, Takashi and
          Obana, Koichiro and
          Arai, Ryuta and
          Kaiho, Yuka and
          Nakamura, Yasuyuki and
          Yokobiki, Takashi and
          Kodaira, Shuichi and
          Takahashi, Narumi and
          Ellwood, Robert and
          Yartsev, Victor and
          Karrenbach, Martin},
  year = {2021},
  month = Feb 
}

@article{Segui2025,
  title = {{DASPack: controlled data compression for distributed acoustic sensing}},
  volume = {244},
  ISSN = {1365-246X},
  DOI = {10.1093/gji/ggaf397},
  number = {1},
  journal = {Geophysical Journal International},
  publisher = {Oxford University Press (OUP)},
  author = {Seguí, A. and Ugalde, A. and Fichtner, A. and Ventosa, S. and Morros, J.R.},
  year = {2025},
  month = Oct 
}

@book{Hartog2017,
  title = {An Introduction to Distributed Optical Fibre Sensors},
  ISBN = {9781315119014},
  DOI = {10.1201/9781315119014},
  publisher = {CRC Press},
  author = {Hartog,  Arthur H.},
  year = {2017},
  month = May 
}

@article{DiMauro2025,
  title = {{Acoustic Tracking of Sperm Whales (Physeter macrocephalus) in the Central Mediterranean Sea Using the NEMO-O$\nu$DE Deep-Sea Observatory}},
  volume = {13},
  ISSN = {2077-1312},
  DOI = {10.3390/jmse13040682},
  number = {4},
  journal = {Journal of Marine Science and Engineering},
  publisher = {MDPI AG},
  author = {Di Mauro,  Letizia Stella and Diego-Tortosa,  Dídac and Sciacca,  Virginia and Riccobene,  Giorgio and Viola,  Salvatore},
  year = {2025},
  month = Mar,
  pages = {682}
}

@article{Lara2020,
  title = {New Insights into the Design and Application of a Passive Acoustic Monitoring System for the Assessment of the Good Environmental Status in Spanish Marine Waters},
  volume = {20},
  ISSN = {1424-8220},
  DOI = {10.3390/s20185353},
  number = {18},
  journal = {Sensors},
  publisher = {MDPI AG},
  author = {Lara,  Guillermo and Miralles,  Ramón and Bou-Cabo,  Manuel and Esteban,  José Antonio and Espinosa,  Víctor},
  year = {2020},
  month = Sep,
  pages = {5353}
}

@article{Rorstadbotnen2023,
  title = {Simultaneous tracking of multiple whales using two fiber-optic cables in the Arctic},
  volume = {10},
  ISSN = {2296-7745},
  DOI = {10.3389/fmars.2023.1130898},
  journal = {Frontiers in Marine Science},
  publisher = {Frontiers Media SA},
  author = {Rørstadbotnen,  Robin André and Eidsvik,  Jo and Bouffaut,  Léa and Landrø,  Martin and Potter,  John and Taweesintananon,  Kittinat and Johansen,  Ståle and Storevik,  Frode and Jacobsen,  Joacim and Schjelderup,  Olaf and Wienecke,  Susann and Johansen,  Tor Arne and Ruud,  Bent Ole and Wuestefeld,  Andreas and Oye,  Volker},
  year = {2023},
  month = Apr 
}

@article{Clark2019,
  title = {Fin whale singing decreases with increased swimming speed},
  volume = {6},
  ISSN = {2054-5703},
  DOI = {10.1098/rsos.180525},
  number = {6},
  journal = {Royal Society Open Science},
  publisher = {The Royal Society},
  author = {Clark,  Christopher W. and Gagnon,  George J. and Frankel,  Adam S.},
  year = {2019},
  month = Jun,
  pages = {180525}
}

@article{wat1987,
  title   = {The 20-{Hz} signals of {Finback Whales} ({Balaenoptera} physalus)},
  volume = {82},
  ISSN = {1520-8524},
  DOI = {10.1121/1.395685},
  number = {6},
  journal = {The Journal of the Acoustical Society of America},
  publisher = {Acoustical Society of America (ASA)},
  author = {Watkins, W. A. and Tyack, P. and Moore, K. E. and Bird, J. E.},
  year = {1987},
  month = dec,
  pages = {1901--1912}
}

@Article{Bouffaut2026pre,
  author       = {Bouffaut, Léa},
  title        = {Distributed Acoustic Sensing for Marine Mammal Monitoring},
  journal      = {ESS Open Archive},
  year         = {2026},
  month        = jun,
  day          = {29},
  doi          = {10.22541/essoar.15005355/v1},
  url          = {https://essopenarchive.org/doi/full/10.22541/essoar.15005355/v1},
  note         = {Preprint, version 1},
}

@article{Horne2025,
  author    = {Horne, Steve and Stork, Anna L. and Stanek, Frantisek},
  journal   = {Frontiers in Marine Science},
  title     = {Observations of fin whales and vessels offshore oregon using fibre optic distributed acoustic sensing},
  year      = {2025},
  issn      = {2296-7745},
  month     = jul,
  volume    = {12},
  doi       = {10.3389/fmars.2025.1603541},
  publisher = {Frontiers Media SA},
}

@article{Woollam2020,
  title = {{HEX: Hyperbolic Event eXtractor},  a Seismic Phase Associator for Highly Active Seismic Regions},
  volume = {91},
  ISSN = {1938-2057},
  DOI = {10.1785/0220200037},
  number = {5},
  journal = {Seismological Research Letters},
  publisher = {Seismological Society of America (SSA)},
  author = {Woollam,  Jack and Rietbrock,  Andreas and Leitloff,  Jens and Hinz,  Stefan},
  year = {2020},
  month = Jul,
  pages = {2769--2778}
}

@article{Havlik2022,
  title = {State of Play in Marine Soundscape Assessments},
  volume = {9},
  ISSN = {2296-7745},
  DOI = {10.3389/fmars.2022.919418},
  journal = {Frontiers in Marine Science},
  publisher = {Frontiers Media SA},
  author = {Havlik,  Michelle-Nicole and Predragovic,  Milica and Duarte,  Carlos Manuel},
  year = {2022},
  month = Jun
}

@article{Merchant2022,
  title = {A decade of underwater noise research in support of the European Marine Strategy Framework Directive},
  volume = {228},
  ISSN = {0964-5691},
  DOI = {10.1016/j.ocecoaman.2022.106299},
  journal = {Ocean \& Coastal Management},
  publisher = {Elsevier BV},
  author = {Merchant,  Nathan D. and Putland,  Rosalyn L. and André,  Michel and Baudin,  Eric and Felli,  Mario and Slabbekoorn,  Hans and Dekeling,  René},
  year = {2022},
  month = Sep,
}

@article{Marques2012,
  title = {Estimating animal population density using passive acoustics},
  volume = {88},
  ISSN = {1469-185X},
  DOI = {10.1111/brv.12001},
  number = {2},
  journal = {Biological Reviews},
  publisher = {Wiley},
  author = {Marques,  Tiago A. and Thomas,  Len and Martin,  Stephen W. and Mellinger,  David K. and Ward,  Jessica A. and Moretti,  David J. and Harris,  Danielle and Tyack,  Peter L.},
  year = {2012},
  month = Nov,
  pages = {287–309}
}

@article{VanParijs2009,
  title = {Management and research applications of real-time and archival passive acoustic sensors over varying temporal and spatial scales},
  volume = {395},
  ISSN = {1616-1599},
  DOI = {10.3354/meps08123},
  journal = {Marine Ecology Progress Series},
  publisher = {Inter-Research Science Center},
  author = {Van Parijs,  SM and Clark,  CW and Sousa-Lima,  RS and Parks,  SE and Rankin,  S and Risch,  D and Van Opzeeland,  IC},
  year = {2009},
  month = Dec,
  pages = {21–36}
}

@article{Baird2025,
  title = {Ocean Space Surveillance and Real-Time Event Characterization Using Distributed Acoustic Sensing on Submarine Networks},
  volume = {96},
  ISSN = {1938-2057},
  DOI = {10.1785/0220240360},
  number = {2A},
  journal = {Seismological Research Letters},
  publisher = {Seismological Society of America (SSA)},
  author = {Baird,  Alan F. and Morten,  Jan Petter and Oye,  Volker and Bjørnstad,  Steinar},
  year = {2025},
  month = Jan,
  pages = {691--705}
}

@Article{Best2022,
  author          = {Best, Paul and Marxer, Ricard and Paris, Sébastien and Glotin, Hervé},
  journal         = {Scientific Reports},
  title           = {Temporal evolution of the Mediterranean fin whale song},
  year            = {2022},
  issn            = {2045-2322},
  month           = aug,
  number          = {1},
  volume          = {12},
  comment-ddietor = {Diferència notes A-B amb freq. 19.96 Hz
8 s window surrounding the prediction peak + BP Btterworth filter (order 3) 10-30 Hz, fs 250 Hz},
  doi             = {10.1038/s41598-022-15379-0},
  publisher       = {Springer Science and Business Media LLC},
}

@article{diego2025,
  author = {Diego-Tortosa, D. and Bonanno, D. and Bou-Cabo, M. and Di Mauro, L. S. and Idrissi, A. and Lara, G. and Riccobene, G. and Sanfilippo, S. and Viola, S.},
  title = {Effective strategies for automatic analysis of acoustic signals in long-term monitoring},
  volume = {13},
  ISSN = {2077-1312},
  DOI = {10.3390/jmse13030454},
  number = {3},
  journal = {Journal of Marine Science and Engineering},
  publisher = {MDPI AG},
  year = {2025},
  month = feb,
}

@article{sciacca2015,
 author = {Sciacca, V. and Caruso, F. and Beranzoli, L. and Chierici, F. and De Domenico, E. and Embriaco, D. and Favali, P. and Giovanetti, G. and Larosa, G. and Marinaro, G. and Papale, E. and Pavan, G. and Pellegrino, C. and Pulvirenti, S. and Simeone, F. and Viola, S. and Riccobene, G.},
  title   = {Annual Acoustic Presence of Fin Whale (\textit{Balaenoptera physalus}) Offshore Eastern Sicily, Central Mediterranean Sea},
  journal = {PLOS ONE},
  year    = {2015},
  volume  = {10},
  number  = {11},
  doi     = {10.1371/journal.pone.0141838}
}

@article{latorre2025,
  title = {{KVP: a multiscale kurtosis approach for seismic phase picking}},
  volume = {241},
  ISSN = {1365-246X},
  DOI = {10.1093/gji/ggaf136},
  number = {3},
  journal = {Geophysical Journal International},
  publisher = {Oxford University Press (OUP)},
  author = {Latorre, H. and Ventosa, S. and Ugalde, A. and Villaseñor, A. and Bartolomé, R. and Ranero, C. R.},
  year = {2025},
  month = Apr,
  pages = {1923--1935}
}

@article{Castellote2011,
  title = {Fin whale (\textit{Balaenoptera physalus}) population identity in the western {Mediterranean Sea}},
  volume = {28},
  ISSN = {1748-7692},
  DOI = {10.1111/j.1748-7692.2011.00491.x},
  number = {2},
  journal = {Marine Mammal Science},
  publisher = {Wiley},
  author = {Castellote,  Manuel and Clark,  Christopher W. and Lammers,  Marc O.},
  year = {2011},
  month = Jul,
  pages = {325--344}
}

@article{Castellote2012,
  title = {Acoustic and behavioural changes by fin whales (Balaenoptera physalus) in response to shipping and airgun noise},
  volume = {147},
  ISSN = {0006-3207},
  DOI = {10.1016/j.biocon.2011.12.021},
  number = {1},
  journal = {Biological Conservation},
  publisher = {Elsevier BV},
  author = {Castellote,  Manuel and Clark,  Christopher W. and Lammers,  Marc O.},
  year = {2012},
  month = Mar,
  pages = {115--122}
}

@inproceedings{Ester1996,
author = {Ester, M. and Kriegel, H.-P. and Sander, J. and Xu, X.},
title = {A density-based algorithm for discovering clusters in large spatial databases with noise},
year = {1996},
publisher = {AAAI Press},
booktitle = {Proceedings of the Second International Conference on Knowledge Discovery and Data Mining},
pages = {226--231},
numpages = {6},
location = {Portland, Oregon},
series = {KDD'96}
}

@misc{SAFE,
  doi = {10.7914/ZGE4-KD84},
  url = {https://www.fdsn.org/networks/detail/7M_2023/},
  author = {Ugalde, A. and Bartolome, R. and Cabieces, R. and Grevemeyer, I. and Villaseñor, A. and Estrada, F. and Palomino, D. and Cubas, M. and Neri, A. and Latorre, H.},
  title = {{SAFE: OBS and DAS acquisition off the coast of Estepona (Spain)}},
  publisher = {International Federation of Digital Seismograph Networks},
  year = {2023}
}

@misc{Tarifa,
  doi = {10.7914/QMCG-2X78},
  url = {https://www.fdsn.org/networks/detail/22_2022/},
  author = {Ugalde, A. and Martín-López, S. and González-Herraez, M.},
  title = {{DAS acquisition in the Strait of Gibraltar}},
  publisher = {International Federation of Digital Seismograph Networks},
  year = {2022}
}

\end{document}